\documentstyle[12pt]{article}

\begin{document}
\begin{titlepage}
\setcounter{page}{1}
\title{\bf Model of the gravitational dipole}
\author{Waldemar Puszkarz\thanks{Electronic address: 
puszkarz@cosm.sc.edu}
\\
\small{\it  Department of Physics and Astronomy,}
\\
\small{\it University of South Carolina,}
\\
\small{\it Columbia, SC 29208}}
\date{\small (October 8, 1997)}
\maketitle
\begin{abstract}
A model of the gravitational dipole is proposed in a close analogy
to that of the global monopole. The physical properties and the
range of validity of the model are examined as is the motion of 
test particles in the dipole background. It is found that the 
metric of the gravitational dipole describes a curved space-time, 
so one would expect it to have a more pronounced effect on the 
motion of the test particles than the spinning cosmic string.
It is indeed so and in the generic case the impact of repulsive 
centrifugal force results in a motion whose orbits when projected 
on the equatorial plane represent unfolding spirals or hyperbolas. 
Only in one special case these projections are straight lines, 
pretty much in a manner observed in the field of the spinning 
cosmic string. Even if  open, the orbits are nevertheless bounded 
in the angular coordinate $\theta$.

\vskip 0.5cm
\noindent
\end{abstract}

\end{titlepage}

\section{Introduction}
Recently, there has been a flurry of activity in studying exotic 
cosmic structures, ranging from strings to monopoles to walls 
[1-4]. The motivation to study these objects comes mainly from 
theories of particle physics, where such structures naturally 
emerge as topological defects which may have arisen during phase 
transitions in the early Universe. Since  they are macroscopic,
to obtain their complete physical description, including global 
properties, one is inevitably lead to incorporate into this picture 
also gravity.\footnote[1] {The history of topological defects in 
gravity begins with a seminal paper of A. Staruszkiewicz in 
{\it Acta Phys. Polon.} {\bf 24}, 734 (1963), where the defect 
analogous to the cosmic string is introduced in the 
1+2- dimensional context.}
On the other hand, one can envisage these or similar structures 
within the general relativity (GR) framework only, especially when 
their field theoretical justification is not  available or hard to 
conceive without more sophisticated models, usually invoking 
additional, hypothetical processes. The case at hand is the cosmic 
spinning string \cite{Maz1} which can be viewed as a generalization 
of a static cosmic string [1, 2]. Unlike the latter, the spinning 
cosmic string carries angular momentum whose source cannot be 
convincingly and decidedly established. The only attempt to 
elucidate the origin of the angular momentum made by Mazur 
\cite{Maz2} relies on a purely microscopic phenomenon of 
superconductivity. This, however, is not the only problem that 
one encounters while studying the objects under discussion. Some 
of them are described by sources which themselves 
are challenged by certain pathologies. A good example to call for 
is that of the global monopole \cite{Bar1}. The energy-momentum 
tensor of that object is provided by a respectable theory. However, 
since the mass of the monopole per radial length is constant, its
density grows infinitely large when we approach the monopole's 
center. Due to this and the physical requirement that the 
monopole's mass be finite one should think of a realistic 
gravitational monopole as a shell rather than a ball. 

The still hypothetical objects mentioned in the previous paragraph 
belong to some sort of extremes: 
the spinning cosmic string in the way it is described by \cite{Maz1}  
extends to infinity, as does the gravitational monopole according 
to the model of \cite{Bar1}. However, these circumstances should 
not discourage us from considering those models
as representing some physical reality to  which they are just the 
first and sometimes crude approximation. In fact, it is usually 
very difficult to find metrics that would give us finer 
descriptions of this reality without the simplicity to be sacrificed, 
be it a finite cosmic string or a global monopole 
of a finite size. Therefore such models are not unfrequently the  
most reasonable compromises between the complexity that a more 
accurate description necessarily entails and the elucidation of 
physical contents which one attempts to achieve without pushing the 
complexity too far. Since it is very often the latter that 
interests us most, it would be rather unwise to discard the models 
that are capable of giving us a glimpse of the contents
at a very little expense of means involved. As we will see, this 
also concerns the model  we aim at presenting here, the 
gravitational dipole, which shares features of the global monopole 
both in its infinite spatial extension and a singularity of its 
internal physical characteristics, in this case the density of 
angular momentum. We do propose to overcome these by suggesting 
some better physical approximation to the strict mathematical 
model in a manner similar to that for the global monopole.
    		
The nature of spin inducing sources in GR is still a subject of
speculations. For example, the Kerr solution \cite{Ker} that we 
will use to derive the metric of the gravitational dipole, believed to
be the metric of the exterior of the rotating black hole, has not
been successfully matched to an interior solution that would be
generated by some material model, like a spinning shell. Some 
progress towards this goal has been achieved \cite{Cru, Isr}, though.   
  
Despite this rather obscure nature of the spinning mechanisms in GR,
or perhaps because of this, one is nevertheless challenged to try to 
understand how the spinning objects can arise and to what physical
phenomena they may lead in either classical or quantum domain of 
physics. The goal of the present paper is to address the latter 
issue. In what follows we propose a model of the 3-dimensional 
gravitational dipole in an attempt to investigate the consequences 
of spinning motions in a general relativistic setting. It is in this 
setting that the gravitational dipole model seems to emphasize the 
role of spin in a more explicit manner than the Kerr black hole. In 
the limit of vanishing angular momentum, the Kerr metric boils 
down to the Schwarzschild metric, both describing curved manifolds, 
while in the very same limit the gravitational dipole space-time 
becomes flat. It is in this context that the difference between the 
two spinning space-times is demonstrated most pronouncedly. The 
content of the gravitational dipole metric is purely rotational, 
without the angular momentum any conceivable gravitational 
interactions are absent here, unlike in the case of the Kerr metric. 
One can therefore hope that through the study of this model one 
would better understand the physical implications of spin in GR in 
three dimensions which may be different than those due to an 
essentially two dimensional spinning source of the string 
\cite{Des}. One would also like to understand how these sources 
can affect the behavior of spin carrying quantum particles.
These investigations will be presented in another paper \cite{Pus}.  

The paper is organized as follows. The next section deals with 
the derivation of the metric for the model under consideration and 
discusses the conditions under which this derivation is valid. It 
also presents the main properties of the model's space-time and 
considers their physical relevance. As explicitly shown in the 
appendix, unlike in the case of the spinning string the space of 
the gravitational dipole is not even locally flat. Because of this, 
the spin induced gravity affects the motion of test particles in a 
manner different from the way the spinning cosmic string imparts 
on them. This motion is more complex than in the field of the string 
as new features characteristic of the centrifugal force are present. 
This is the main observation of the section that follows the 
derivation of the metric and precedes the conclusions that 
summarize our findings. 
																		
\section{The metric of the gravitational dipole and assumptions that 
underline it}

We will derive here the metric of the gravitational 
dipole and discuss the circumstances under which it can be valid as 
a consistent description of some sort of physical reality. To this 
end, we start from the Kerr metric in the Boyer-Lindquist coordinates 
\cite{Boy}
$$
\begin{array}{r} ds^2=\frac{\Delta}{\Sigma^2}
(dt-a\sin^2\theta d\phi)^2-\frac{\sin^2\theta}{\Sigma^2}
[(r^2+a^2)d\phi-ad\phi]^2-\frac{\Sigma^2}{\Delta}-
\Sigma^2d\theta^2=
\\ 
\frac{1}{\Sigma^2} (\Delta-a^2\sin^2\theta)dt^2+\frac{2\sin^2\theta}
{\Sigma^2}[a(r^2+a^2)-a\Delta]dtd\phi-\frac{\Sigma^2}{\Delta}dr^2-
\Sigma^2d\theta^2+\\ 
\frac{\sin^2\theta}{\Sigma^2}
[a^2\Delta\sin^2\theta-(r^2+a^2)^2]d\phi^2 ,\end{array} \eqno (1)
$$
where $\Sigma^2=r^2+a^2\cos^2\theta$, $\Delta=r^2-2Mkr+a^2$, and 
$a=\frac{J}{M}$. Here $J$ denotes the angular momentum of the 
rotating black hole, $M$ its mass, and $k=\frac{G}{c^2}$. We will 
assume that $J=jr$, $M=mr$, $j$ and $m$ being some constants and 
that terms proportional to and of higher order in $p=\frac{a}{r}$ 
can be neglected in a certain sensible approximation that we want 
to work out. This amounts to saying that $p\ll1$ in the 
approximation intended. We will later substantiate this approach 
in greater detail. We will also assume that $km\ll1$, $kj\ll1$ and 
because of this the terms $kmp$ (and obviously also $kmp^2$) will 
not be taken into account in a series expansion of the metric (1) 
in parameters $km$, $kj$, and $p$ that seem to be the most natural 
ones for this sort of expansion.

Now, from (1) we obtain up to $O(p^2)$ but with $O(kmp^2)$ already 
discarded that
$$
 \begin{array}{*{3}{l}} 
g_{tt}=1-2km,& -g_{rr}=1+2km-p^2\sin^2\theta,& 
-g_{\theta\theta}= r^2(1+p^2\cos^2\theta),\\            
-g_{\phi\phi}=r^2(1+p^2)\sin^2\theta,&
-g_{t\phi}=2kj\sin^2\theta.  \end{array} \eqno (2)
$$
This leads us to the metric:
$$
\begin{array}{r} ds^2= (1-2km)(dt-2kj\sin^2\theta 
d\phi)^2-(1+2km)(1-\frac {p^2\sin^2\theta}{1+2km})dr^2  
-\\r^2[(1+p^2\cos^2\theta)d\theta^2 +(1+p^2)\sin^2\theta d\phi^2]+ 
O(k^2j^2)d\phi^2. \end{array}  \eqno (3)
$$
Neglecting terms $O(k^2j^2)$ and $O(p^2)$ gives
$$
ds^2=(1-2km)(dt-2kj\sin^2\theta d\phi)^2-(1+2km)dr^2-
r^2d\theta^2-r^2\sin^\theta d\phi^2 . \eqno (4)
$$
Rescaling (4) by $(1-2km)^{-1}$ and changing the coordinates via
$d\rho=(1+2km)dr$ brings us to
$$
ds^2=(dt-2kj\sin^2\theta d\phi)^2-d\rho^2-(1-2km)\rho^2
(d\theta^2+\sin^2\theta d\phi^2),\eqno (5)
$$
where we used $(1-2km)^{-1}=1+2km$ to be consistent with our 
assumption $km\ll1$.
Therefore we arrived at the metric
$$
ds^2=(dt-A\sin^2\theta d\phi)^2-d\rho^2-B\rho^2
(d\theta^2+\sin^2\theta d\phi^2),
\eqno (6)
$$
where $A=2kj$ and $B=1-2km$.
Since $B\ne1$ represents a purely monopole contribution to the 
metric (6) as seen from the metric of the global monopole \cite{Bar1},
$$
ds^2=dt^2-d\rho^2-b^2\rho^2(d\theta^2+\sin^2\theta d\phi^2),
$$
and we are interested in the purely dipole part of it, we will put $B$ 
equal $1$ in order to focus on the dipole content of our derivation.
This is tantamount to saying that
$$
ds^2=(dt-A\sin^2\theta d\phi)^2-dr^2-r^2
(d\theta^2+\sin^2\theta d\phi^2)
\eqno (7)
$$
is the metric of the gravitational dipole under the assumptions 
specified earlier.
									
We will now investigate if these assumptions are self-consistent and
what is the physical nature of the object that produces the 
gravitational field described by (7). The source of the field is 
the momentum-energy tensor whose components are proportional to the
components of the Einstein tensor and the latter are worked out in 
the appendix. Because of their complexity it is rather hard to propose 
a field-theoretical model, or what amounts to the same, some matter 
Lagrangian that would lead to this energy-momentum tensor. This,
however, as argued in the introduction, should not be 
considered a major obstruction.
The metric (7), due to the assumption that the angular momentum per
unit radial length is constant, describes an object whose total 
angular momentum is infinite. To make this situation look more 
physical one can think of (7) as a good approximation to the metric
of an exterior of a shell of thickness $d$ in its vicinity, 
a distance $r\gg a$ from the center, which is required to satisfy 
the condition $p\ll1$. Then the angular momentum is finite and 
proportional to $d$. We will soon show that under reasonable and
physically natural circumstances one can make $a=1$ and since 
making $a=1$ coincides with  imposing on $r$ the condition 
$r\ge\frac{1}{(km)^{1/2}}$ the discussed condition $p\ll1$ is not 
restrictive at all. As a matter of fact, because $km\ll1$, $p\ll1$ is 
easily met. Why it should be a shell rather than a ball can be 
justified by equilibrium considerations. For the equilibrium to
take place the centripetal force present in any rotating system 
should not overcome the gravity force, that is for a test particle
of mass $\mu$ there should be 
$\mu\omega^2(r)r\le\frac{kM(r)\mu}{r^2}$,
where $\omega(r)$ and $M(r)$ are the angular velocity and the mass
of the matter that generate our metric. This means that
$$
\omega^2(r)\le \frac{kM}{r^3}={km}{r^2}.  \eqno (8) 
$$
On the other hand, from the definition of the angular momentum
$J=\int r^2\omega(r)\,dM(r)$ one obtains
$$
\frac{dJ}{dr}=j=const=r^2\omega(r)\frac{dM}{dr}=r^2\omega(r)m 
\eqno (9)
$$
according to the assumptions employed.
Eq. (9) requires $\omega(r)\sim \frac{1}{r^2}$, thus leading to
a sigularity with $r\rightarrow 0$. It is the main reason why we
want a shell as  our model for the metric of the gravitational
dipole. The situation here is somewhat similar to the mass
distribution in the global monopole, where $\frac{dM}{dr}$ 
being constant implies $\rho(r)\sim \frac{1}{r^2}$, which holds
also for the model of the gravitational dipole under study.  
Obviously, $\omega$ cannot grow to infinity without 
destroying the stability of the whole configuration. It is 
Eq. (8) that demands for $\omega(r)=\frac{C}{r^2}$ a minimum 
radius $r_0$ such that $\frac{C^2}{r_0^4}=\frac{km}{r_0^2}$ or
$r_0=\sqrt{\frac{C^2}{km}}$. By choosing $C=1$, which can always
be accomplished by rescaling the length, we see that the balance
of the configuration is maintained as long as its inner radius is
greater than $\frac{1}{(km)^{1/2}}$. This choice of $C$ leads via 
(9) to $a=\frac{j}{m}=1$.
  
A few remarks on physical properties of the introduced space-time 
seems to be in order here. First of all, the metric (7) as an 
axially-symmetric one possesses two Killing vectors $\partial_{t}$ 
and $\partial_{\phi}$. In the region $r<A\sin\theta$, the latter, 
which has closed orbits, becomes timelike. Any observer following 
these orbits would experience causality violation. Fortunately, in 
the regime in which the metric of the gravitational dipole is valid 
(that is $km\ll 1$ and so also $kj\ll 1$ for  $a=\frac{j}{m}=1$)
this region lies inside the shell discussed above and its physical 
relevance is only apparent. As long as one stays outside of this 
shell, one can dismiss this issue.

As shown in \cite{Har} the spinning cosmic string indroduces the 
time delay between the arrival time of two particles moving in 
opposite directions that follow an arbitrary closed trajectory 
around the string. One can rightly expect a similar effect in the 
case under study. The only difference between these two physical 
situations is that in the metric of the gravitational dipole
this delay depends on the angle $\theta$. For light rays describing 
circular paths with $\phi=\Omega t$, $r=R=$const, and $\theta=$const, 
the eikonal equation $ds^{2}=0$ implies 
$\Omega_{+}=\frac{1}{R-A\sin^2\theta}$ and 
$\Omega_{-}=-\frac{1}{R+A\sin^2\theta}$ with the $\Omega_{+}$ 
assigned to the ray moving along with the direction of rotation. 
Assuming $R>A\sin^2\theta$, which, as argued earlier, is always valid 
for our model, one obtains for the time delay
$$
T=2\pi\left[\frac{1}{|\Omega_{-}|}-\frac{1}{\Omega_{+}}\right] \eqno (10)
$$

The last comment concerns the very name we chose for the model presented.
As is well-known, the gravitational charges are only alike, thus 
preventing the existence of ``electric'' gravitational dipoles. However, 
the dipole we introduced in this paper is of ``magnetic'' nature. Since 
in GR every form of energy contributes  to the gravitational field, 
in the case under consideration it is the rotational (``magnetic'') 
energy of the dipole that does so.
As opposed to the spinning cosmic string that carries a nonzero 
flux of the ``magnetic'' field, but does not produce a detectable 
field itself \footnote[2] {For this reason the spinning cosmic string 
could also be called the gravitational fluxon.}, the gravitational 
dipole creates such a field as manifested by the curvature of its 
space-time.

\section{The motion of test particles}

As observed in the previous section, the metric of the gravitational
dipole possesses two Killing vectors, $\partial_{t}$ and 
$\partial_{\phi}$ that can be used to find integrals of motion of
test particles in the space-time of the dipole. The integrals 
themselves are useful in simplifying the equations of motion for
the particles by casting them into a dynamical system. To employ 
this strategy let us recall that 
$$
(\partial_{\alpha}|\dot{\xi})={\rm const}=\frac{1}{\mu_0}P_{\alpha} 
\eqno (1)
$$ 
along the geodesic $\xi$, where $\partial_{\alpha}$ is a Killing 
vector, $\dot{\xi}$ is a vector tangent to the geodesic, and 
$P_{\alpha}$ is a momentum conjugate to the coordinate $x^{\alpha}$,
chosen so that $\partial_{\alpha}$ is tangent to it. Eq. (1) is 
valid also for a zero-mass particles, i.e., when $\mu_0=0$ except 
for its last equality. The brackets in (1) denote the scalar product 
of two vectors separated by a vertical dash.

Eq. (1) when applied to an axially-symmetric metric yields 
$$
g_{tt}\dot{t}+g_{t\phi}\dot{\phi}=-\frac{E}{\mu_{0}c^2}=-\gamma, 
\eqno (2a)
$$ 
$$ g_{tt}\dot{t}+g_{\phi\phi}\dot{\phi}=\frac{l}{\mu_{0}c}=
\lambda, \eqno (2b) 
$$
where $\gamma$ and $\lambda$ are constants related to the energy 
$E$ and the azimuthal angular momentum $l$ of the test particle. 
When used for our metric Eqs. (2) lead to 
$$ 
\dot{t}=\frac{\lambda A-\gamma X}{r^2}, \eqno (3a) 
$$ 
$$ 
\dot{\phi}=\frac{\lambda- A \gamma}{r^2}, \eqno (3b) 
$$ 
where
$X=r^2-A^2\sin^2\theta$. The remaining equations of motion can 
be obtained from the Hamilton-Jacobi equation 
$$ g^{\mu\nu}\frac{\partial S}{\partial
x^{\mu}}\frac{\partial S}{\partial x^{\nu}}-\mu_{0}^2c^2=0. 
\eqno (4) 
$$
In this formalism $P_{\mu}=\frac{\partial S}{\partial x_{\mu}}$ 
and since we have already established that $P_{t}=-\mu_{0}c\gamma$ 
and $P_{\phi}=\mu_{0}c\lambda$ one easily arrives at 
$$
\left(1-\frac{A^2\sin^2\theta}{r^2}\right)P^2_{t}-\frac{1}{r^2}
P^2_{\theta}-P^2_{r}-\frac{2A}{r^2}P_{t}P_{\phi}-\frac{1}
{r^2sin^\theta}P^2_{\phi}-\mu^2_{0}c^2=0 
$$ and then at 
$$
-\left(\frac{P^2_{\theta}}{\mu^2_{0}c^2} +\frac{\lambda^2}
{\sin^2\theta}+\gamma^2A^2\sin^2\theta\right) + r^2 \left[
-\frac{P^2_{r}}{\mu^2_{0}c^2}+(\gamma^2-1)+ \frac{2A\lambda 
\gamma}{r^2} \right]=0 .\eqno (6) 
$$ 

By introducing a separation constant $K$ such that 
$$
\frac{P^2_{\theta}}{\mu^2_{0}c^2} 
+\frac{\lambda^2}{\sin^2\theta}+\gamma^2A^2\sin^2\theta=K^2
 \eqno (7) 
$$ 
one obtains 
$$ \frac{P^2_{r}}{\mu^2_{0}c^2}=(\gamma^2-1)+
\frac{K^2-2A\lambda \gamma}{r^2}. 
\eqno (8) 
$$ 
With the substitution $\dot{x}^{\alpha}=\frac{P_{\alpha}}
{\mu_{0}c}$ Eqs. (7) and (8) can be reduced to the following 
dynamical system for $r$ and $\theta$ 
$$ 
\dot{r}=\left(\gamma^2-1-\frac{K^2-2A\lambda\gamma}{r^2}\right)
^{1/2}, \eqno (9a) 
$$ 
$$\dot{\theta}=\left(K^2- \frac{\lambda^2}{\sin^2\theta}-
\gamma^2A^2\sin^2\theta\right)^{1/2}. \eqno (9b)
$$ 

Making the substitution $z=\cos\theta$, this last equation can be
transformed into the energy conservation equation for the 
one-dimensional motion of unit mass particle in the potential 
$V(z)$, or explicitly
$$
\frac{\dot{z}^2}{2}+V(z)=E \eqno(10)
$$
with $E=\frac{1}{2}(K^2-\lambda^2-\gamma^2A^2)$ and $V(z)=az^2+bz^4$, 
where $a=\frac{1}{2}(K^2-2\gamma^2A^2)$ and $b=\frac{1}{2}\gamma^2A^2$. 
One more substitution, $x=z^2$, brings us to the equation
$$
\dot{x}^2=8x(E-ax-bx^2) \eqno(11)
$$
which when supplemented with the constraints $\dot{x}^2\ge0$ and 
$x\ge0$ describes the physical motion in the coordinate $\theta$.

Depending on the values of $E$ and $a$, there are different ranges 
of $\theta$ in which this motion takes place. Let us first consider 
the case of $a\ge0$. As can be easily seen, the motion is possible 
only for $E\ge0$, $E=0$ representing a singular situation with the 
orbit constrained to the equatorial plane $\theta=\frac{\pi}{2}$. 
For $E<0$ both $x_{-}$ and $x_{+}$ are negative and upon consulting 
with (12) one sees that the motion is impossible ($\dot{x}^2<0$). 
When the ``total energy'' $E$ is positive, the trajectory is 
bounded between this plane and the plane determined by the equation 
$\theta_{-}=\arccos\sqrt{x_{-}}$, where $x_{-}=-\frac{a-\sqrt{a^2+4bE}}
{2b}$. For $a<0$ there are more options and only if 
$E<E_{min}=-\frac{a^2}{4b}$ no motion takes place. For $E=E_{min}$ 
there exists a stationary orbit in the plane 
$\theta=\arccos\sqrt{x_{2}}$ with $x_{2}=-\frac{a}{2b}$. This 
corresponds to the previous situation of $E=0$. When the 
``total energy'' exceeds $E_{min}$ there exists an orbit bounded 
by the planes $\theta_{-}=\arccos\sqrt{x_{-}}$, and 
$\theta_{+}=\arccos\sqrt{\max[0, x_{+}]}$, 
where $x_{+}=-\frac{a+\sqrt{a^2+4bE}}{2b}$. 
This also includes the case $E=0$ for which $\theta_{+}=\frac{\pi}{2}$.

If both $a$ and $E$ are negative the solution to (11) can be expressed 
in terms of elliptic functions. To see this let us rewrite this 
equation as 
$$
\dot{x}^2=-8bx(x-x_{+})(x-x_{-}), \eqno (12)
$$
where $x_{\pm}$ are defined above.
Upon the introduction of $k^2=\frac{x_{+}}{x_{-}}$ and 
$y$ such that $x=x_{+}y^2$, the last equation becomes
$$
\dot{y}^2=-2bx_{-}(1-y^2)(1-k^2y^2), \eqno (13)
$$
Through the substitution $u=i\sqrt{2bx_{+}}t$ one can bring (13) to
the standard form of the differential equation for elliptic 
functions
$$
\left(\frac{dy}{du}\right)^2= (1-y^2)(1-k^2y^2),
$$
whose solution is $y(u)=sn(u+\delta)$ with $\delta$ being some constant.
Finally, the solution to (9b) is found to be
$$
\theta(t)=\arccos\left[\sqrt{x_{+}}
\left|{\rm sn}\left(\sqrt{2bx_{-}}it+\delta\right)
\right|\right]. \eqno (14a)
$$

If $E=0$ and $a<0$ then $x_{+}=0$ and (12) reduces to 
$$
\left(\frac{dy}{du}\right)^2= 2bx_{+}y^2(1-y^2),
$$
where $y^2=\frac{x}{x_{-}}$. Upon the substitutions 
$u=\sqrt{2bx_{-}}t$ and $1-y^2=w^2$ the last equation boils down to
$$
\left(\frac{dw}{du}\right)^2=(1-w^2)^2,
$$
solution to which is given by $w=\pm\tanh (u+c)$, $c$ being some constant.
Eventually, we find that 
$$
\theta(t)=\arccos\left[\sqrt{x_{-}}\left|1-
\tanh^2\left(c+\sqrt{2bx_{-}}t\right)\right|\right]. \eqno (14b)
$$

For $E$ positive and independently of $a$, the trajectory is bounded 
between the equatorial plane and the plane 
$\theta_{-}=\arccos\left(-\frac{a-\sqrt{a^2+4bE}}{2b}\right)^{1/2}$. 
This case can be reduced to the previous one
by changing $x_{+}$, which is always negative, to $ -|x_{+}|$, and 
replacing $x$ by $y$ via $x=y-|x_{+}|$. In doing so one ends up 
with 
$$
\dot{y}^2=-8by\left(y-(x_{-}+|x_{+}|)\right)
$$
that looks exactly like the equation leading to (14b), except
that $x_{-}$ gets replaced by $x_{-}+|x_{+}|$. Obviously,
the solution to (9b) for such conditions is
$$
\theta(t)=\arccos\sqrt{(x_{-}+|x_{+}|)
\left(1-\tanh^2\left(c+\sqrt{2b(x_{-}+|x_{+}|)}t\right)
\right)^2 -|x_{+}|}. \eqno (14c)
$$

However, the fact that the motion is bounded within some range of
$\theta$ does not necessarily mean that its orbits are closed.
As a matter of fact, we will now demonstrate that the orbits
of test particles are open. To this end let us find the 
trajectory of the test particle as a function  $r(\phi)$. 
By combining (3b) and (9a) one obtains 
$$
d\phi=\frac{Cdr}{r^2(\alpha^2-\beta^2/r^2)}, \eqno (15)
$$
where $C=\lambda-\gamma A$, $\alpha^2= \gamma^2-1>0$, and 
$\beta^2=K^2-2A\lambda\gamma$.
Let us now consider three cases of $\beta^2>0$, $\beta^2=0$, and 
$\beta^2<0$. They will lead to different classes of trajectories. 
In the first case, changing variables $\frac{1}{r}=u$ subsequently 
$u=\frac{\beta}{\alpha}\cos\chi$  brings one to the solution
$$
\frac{1}{r(\phi)}=\frac{\alpha}{|\beta|}\cos
\left[|\beta|(\phi+B)/C\right], 
\eqno (16)
$$
where $B$ is a constant of integration. The case $\beta^2=0$ is 
the easiest to work out. Through only one change of variables 
$\frac{1}{r}=u$, it immediately leads to 
$$
r(\phi)=-\frac{C}{\alpha(\phi+B)}.  \eqno (17)
$$
In the last case, the change $u=\frac{\alpha}{|\beta|}\sinh\chi$ 
enables one to arrive upon integration at
$$
\frac{1}{r(\phi)}=\frac{\alpha}{|\beta|}\sinh
\left[-|\beta|(\phi+B)/C\right]. 
\eqno (18)
$$
As seen from (3b), $C>0$ ensues that $\phi$ increases with time. 
However, the sign of $C$ has no bearing on the character of motion, 
all it does is to change its direction. The trajectories in the 
last two cases describe unfolding spirals which asymptotically 
approach infinity in the limit $\phi\rightarrow B$. For $C>0$ this 
is achieved through increasing values of $\phi$ that can run from 
$-\infty$ while for $C<0$ the orbit develops towards decreasing 
values of $\phi$ with $\phi$ running from $+\infty$. That it is
so is determined by the condition $r(\phi)\ge0$. In the first case 
the trajectories are hyperbolas unless $\beta=\pm C$ which 
corresponds to $E=0$ and results in a motion along a straight line 
in the equatorial plane. Both the hyperbolas and the straight line 
make the closest approach to the center a distance 
$\frac{|\beta|}{\alpha}$ from it. The asymptotes of the
hyperbolas are defined by the equation 
$|\beta|(\phi+B)/C=\pm\frac{\pi}{2}$.
When the motion is projected onto the equatorial plane, 
the asymptotes divide this plane into four sectors with the particle 
moving in the one that satisfies the condition 
$\cos \left[|\beta|(\phi+B)/C\right]>0$.

The equations of motion for particles of zero rest mass can be
worked out in a similar manner. In fact, if the mass term is 
dropped in (5) the only change it brings about is in Eqs. (6),
(8), and (9a), where $\gamma^2-1$ is replaced by $\gamma^2$ 
which represents the square of the total energy of the 
particle. The way this change affects the presented solutions
is really insignificant.

\section{Conclusions}

We presented the model of the gravitational dipole in a yet another 
attempt to investigate the impact of spinning sources on gravity in 
a general relativistic framework. The model studied has some 
advantages in this respect over the Kerr black hole as its 
gravitational field is of pure rotational origin, thus allowing 
one to examine this problem in its disentanglement from the main 
gravity source, the mass. We showed that the model has sensible 
properties if it is thought of as an approximation to a rotating 
shell. The space-time of the dipole is curved, therefore it should 
not come as a surprise that the way gravity associated with 
the curvature affects the motion of test particles is more pronounced 
and complex than that of the spinning cosmic string. Unlike for the 
spinning defect-free string, the trajectories of test particles in 
the field of the dipole when projected on the equatorial plane are 
not straight lines anymore, the only exception being the motion in 
the plane itself. All trajectories are open and 
bounded in the $\theta$ coordinate.

\section*{Acknowledgments}

I would like to thank Professor Pawel O. Mazur for suggesting
this subject to study and his comments. 
This work was partially supported by the NSF grant 
No. 13020 F167 and the ONR grant R\&T No. 3124141.

\section*{Appendix}

The only non-zero covariant components of the Einstein tensor for 
the metric of the gravitational dipole are as follows

$$
G_{tt}=-\frac{3a^2\cos^2\theta}{r^4},
$$
$$
G_{rr}=\frac{a^2\cos^2\theta}{r^4},
$$
$$
G_{\theta\theta}=\frac{a^2\cos^2\theta}{r^2},
$$
$$
G_{\phi\phi}=-\frac{a^2(5 r^2 \cos^2\theta-3r^2 
\cos^4\theta+3a^2\cos^2\theta-6a^2\cos^4\theta +3a^2\cos^6\theta)}{r^4},
$$
$$
G_{\phi t}=G_{t\phi}=-\frac{a(r^2-r^2\cos^2-
3a^2\cos^2\theta+3a^2\cos^4\theta)}{r^4}.
$$
The Ricci scalar $R$ is
$$
R=\frac{2a^2\cos^2\theta}{r^4}.
$$
Clearly, the space-time of the gravitational dipole is curved.

\bigskip

\end{document}